\documentclass{aa}
\usepackage{graphicx}
\begin{document}
   \title{Proper motion surveys of the young open clusters Alpha
   Persei and the Pleiades}
   \titlerunning{}
   \subtitle{}

   \author{N.R. Deacon
          \and
          N. C. Hambly
          }
   \authorrunning{Deacon \& Hambly}
   \institute{Institute for Astronomy, University of
              Edinburgh, Blackford Hill, Edinburgh EH9 3HJ\\
              \email{nd@roe.ac.uk,n.hambly@roe.ac.uk}
             }

   \date{Received ---; accepted ---}

   \abstract{
In this paper we present surveys of two open clusters using photometry
and accurate astrometry from the SuperCOSMOS microdensitometer. These
use plates taken by the Palomar Oschin Schmidt Telescope giving a wide
field ($5^{\circ}$ from the cluster centre in both cases),
accurate positions and a long time baseline for the proper
motions. Distribution functions are fitted to  proper motion vector
point diagrams yeilding formal membership probabilities. Luminosity and
mass functions are then produced along with a catalogue of high
probability members. Background star contamination
limited the depth of the study of Alpha Per to $R=18$. Due to this the
mass function found for this cluster could only be fitted with a power
law ($\xi(m) = m^{-\alpha}$) with $\alpha=0.86^{+0.14}_{-0.19}$. However with the
better seperation of the Pleiades' cluster proper motion from the
field population results were obtained down to $R=21$. As the mass function
produced for this cluster extends to lower masses it is possible to
see the gradient becoming increasingly shallow. This mass function is
well fitted by a log normal distribution.\\

   }

   \maketitle
%
\section{Introduction}
The young open cluster Alpha Per lies at a distance of 183 pc
($(M-m)_{o}=6.31$, van Leeuwen 1999) with an age of 90 Myr (Stauffer, Barrado
\& Bouvier 1999). We assume in this paper that  $A_{R}=0.23$(O'Dell, Hendry \&
Collier Cameron 1994). It has been reasonably well studied in the past using various
techniques. Prosser's catalogue (Prosser 1993) contains a
comprehensive list of candidate member stars identified by photometry, spectroscopy and
radial velocity measurements. The faintest of these has
$V \approx 21.8$($M_{V} \approx 15.5$). These candidate stars have been
supplemented by those identified in similar studies by Zapatero~Osorio
et al.~(1996), Stauffer et al.~(1999)
and Rebolo et al.~(1992). In addition Prosser, Randich \& Simon~(1998)
identified candidate stars which were optical counterparts to X-ray
sources within the cluster.\\
 An early proper motion study by Heckmann,
Dieckvoss \& Kox~(1956) identified stars with proper motions
consistent with cluster membership. However this study only went down
to $V \approx 12$ and did not calculate formal membership
probabilities. Prosser (1992) also used stellar proper motions as an
initial selection technique. Although this study was deeper than the
earlier one ($V=18.8$) it also did not calculate formal membership
probabilities. \\
Most recently Barrado~y~Navascu\'{e}s et al~(2001) produced a deep CCD
survey of the cluster. This identified several candidate low mass
stars and Brown Dwarfs down to $I_{c}=22$. They then used this data to
produce a mass function. The mass function was fitted with a power law
whose slope was found to be $\alpha = 0.59$.\\   
The Pleiades is the best studied open cluster, here we assume
$(M-m)_{o}=5.4$ (van Leeuwen 1999) and $A_{R}=0.15$ (O'Dell, Hendry \&
Collier Cameron 1994). In recent years several
deep CCD surveys of the cluster such as Dobbie et al (2002)
and Moraux et al (2003) have produced candidates down to
$0.03 M_{\odot}$. These have been complimented by proper motion
studies such as those of Hambly et al~(1999) and Moraux, Bouvier and
Stauffer (2002). However the only wide field, high precision proper motion
survey is that of Hambly, Hawkins and Jameson~(1991). Adams et al (2001)
used 2MASS data along with proper motions to produce a wide field
study. This contained many candidate stars outside the tidal radius of
the cluster showing background star contamination is a problem for
many techniques.     
%
   \begin{figure}
   \resizebox{\hsize}{!}{\includegraphics{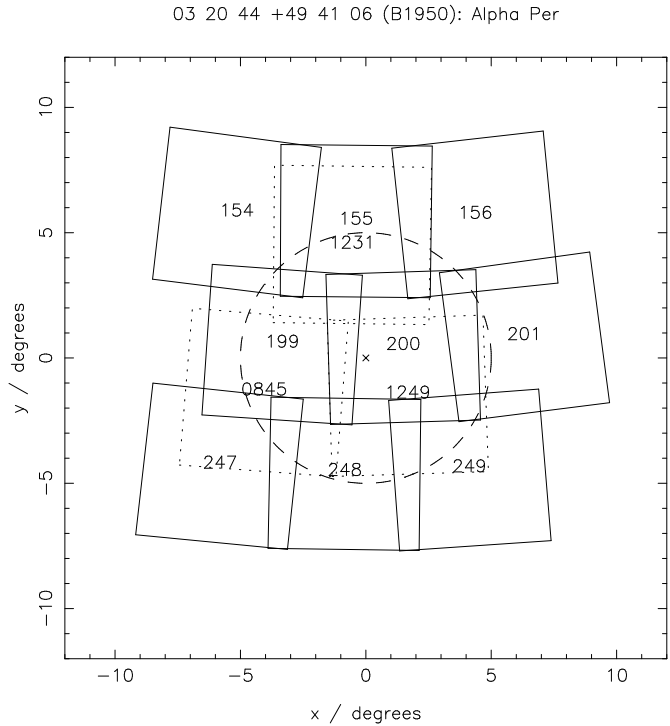}}
      \caption{A plot showing the sky coverage of the Alpha Per
      study. The solid line represent the later POSSII fields while
      the dotted lines are the earlier POSSI fields. The dashed
      circles marks out a five degree radius from the cluster
      centre (marked by an x).}
         \label{applot}

   \end{figure}
%
   \begin{figure}
   \resizebox{\hsize}{!}{\includegraphics{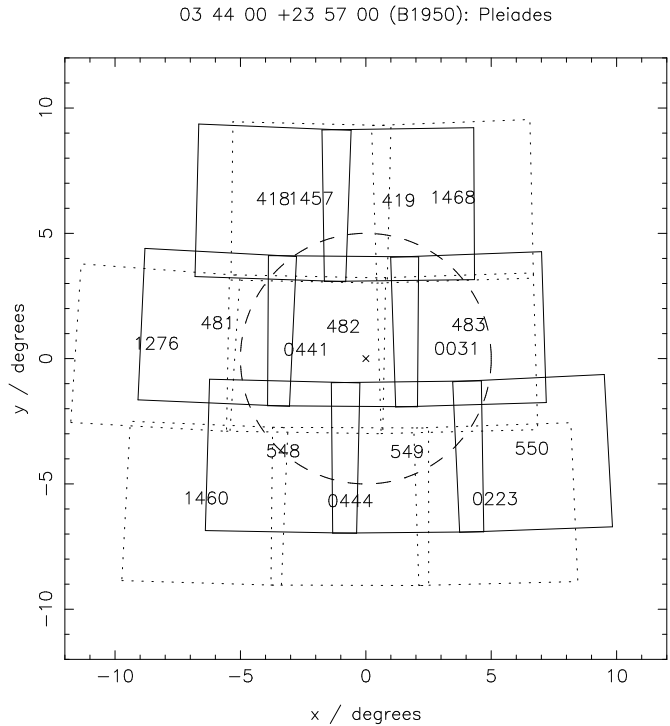}}
      \caption{A plot showing the sky coverage of the Pleiades
      study. The solid line represent the later POSSII fields while
      the dotted lines are the earlier POSSI fields. The dashed
      circles marks out a five degree radius from the cluster
      centre (marked by an x).}
         \label{plplot}

   \end{figure}
\section{Observational data and reduction}
\subsection{Obseravtions}
The SuperCOSMOS facility at the Royal Observatory Edinburgh has
(using plates from the United Kingdom Schmidt Telescope) produced
complete southern sky surveys in $B_{J}$, $R$ and $I$ with an
additional second epoch $R$ survey. These surveys are now publicly
available (Hambly et al 2001). The scanning program has now moved
on to the northern hemisphere, using film and glass copies of plates taken by the Oschin Schmidt Telescope on
Mount Palomar, California. These data will soon be publicly
available. The declination of Alpha Per ($\delta \approx 49^{\circ}$)
and the Pleiades ($\delta \approx 24^{\circ}$) means
this prerelease northern hemisphere data must be used. Details of these plates are given in
Table~\ref{plates} for Alpha Per and Table~\ref{PLplates} for the
Pleiades. Unfortunately due to saturation caused by the bright core
stars of the cluster and a satellite track, 3.1\% (2.3 sq deg) of the
Pleiades survey area was unusable. The region of the CO cloud near Merope was
used in the proper motion survey, it is assumed that the increased
reddening in this region will not affect the results. The areas covered by both surveys are shown
in figures~\ref{applot} and~\ref{plplot}.
\subsection{Sample selection}
The sample of stars used was selected to maximise completeness and
reliability (e.g. by minimising astrometric errors). Highly elliptical objects, non-stellar objects and
poor quality images near bright stars were all removed. Deblended images
were included. This is because Alpha Per (lying at a galactic
latitude of $-7^{\circ}$) is in a crowded field meaning many stellar
images will be merged with others on the plate. Unfortunately the
deblending algorithm is rather crude and  this can lead to increased astrometric
and photometric errors. It is necessary to calculate the astrometric
errors caused by the deblending alogorithm before deciding whether to
include deblended objects. 
%
   \begin{figure}
   \resizebox{\hsize}{!}{\includegraphics{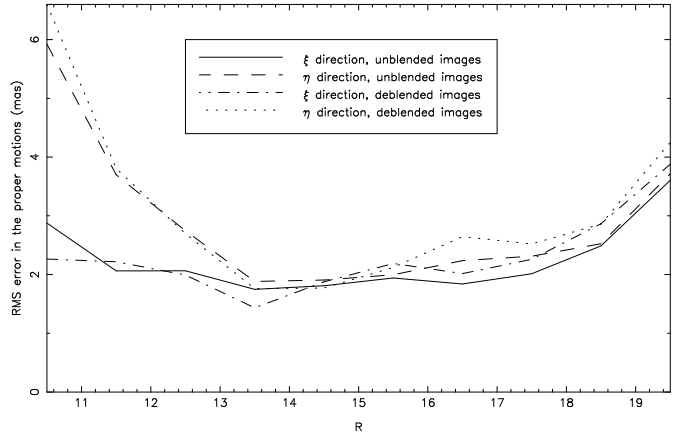}}
      \caption{A plot showing how the errors in the proper motion vary
      with magnitude. The larger errors for bright stars in the $\eta$
      axis are an artifact of the the plate measuring process due to
      saturated images.}
         \label{errplot}

   \end{figure}
\subsection{Errors in the proper motions}
In the Alpha Per data set there were several overlap regions between the plates used in this
study. Hence objects in this region would have two different
measurements of their position. These were used to calculate the RMS
error on the proper motions. The errors were calculated in both axes
($\xi$ and $\eta$) and for objects that had been deblended and those
that had not. Figure~\ref{errplot} shows how these errors vary with
magnitude. Notice the higher errors at the bright end and the faint
end. The overlap regions in the Pleiades were too small
to be used for a similar calculation. Hence a small selection of faint
blue stars was selected. These are expected to have negligible proper
motions, so any measured proper motion would be the result of random
errors. This yielded an error estimates of $\sigma_{x}=6.2$ mas/yr and
$\sigma_{y}=6.3$ mas/yr . It should be noted that these errors are
high because the stars used are faint, Figure~\ref{errplot} shows that
for fainter stars errors are higher. When a study was done
using candidate stars from other studies for Alpha Per it was found
that 172 out of 323 were merged with other images. As it has been
shown that there is no significant increase in the errors for
deblended stars it was decided to include these to maximise completeness. 
   \begin{table*}
\begin{center}
      \caption{Schmidt plates used in the Alpha Per study.}
\begin{tabular}{cccccccccc}
   \hline
Plate no.&Material&Emulsion&Filter&Field no.&Epoch&Exposure&\multicolumn{2}{c}{Plate Centre}&Notes\\
&&&&&&Time&RA&Dec&\\
&&&&&&(minutes)&\multicolumn{2}{c}{B1950}&\\
\hline
3623&3mm glass&IIIaJ&GG385&155&1990.796&60&3 18 00&+55 00 00&$B_{J}$ Plate\\
2730&3mm glass&IIIaJ&GG385&199&1989.697&60&3 00 00&+50 00 00&''\\
2738&3mm glass&IIIaJ&GG385&200&1989.700&60&3 30 00&+50 00 00&''\\
2718&3mm glass&IIIaJ&GG385&248&1989.689&60&3 16 00&+45 00 00&''\\
2843&3mm glass&IIIaF&RG610&155&1989.774&60&3 18 00&+55 00 00&2nd epoch R\\
2822&3mm glass&IIIaF&RG610&199&1989.767&60&3 00 00&+50 00 00&''\\
2832&3mm glass&IIIaF&RG610&200&1989.769&60&3 30 00&+50 00 00&''\\
2817&3mm glass&IIIaF&RG610&248&1989.763&60&3 16 00&+45 00 00&''\\
5498&film&IVN&RG9&155&1993.772&60&3 18 00&+55 00 00&I Plate\\
6569&film&IVN&RG9&199&1995.875&60&3 00 00&+50 00 00&''\\
4388&film&IVN&RG9&200&1991.952&60&3 30 00&+50 00 00&''\\
6057&film&IVN&RG9&248&1994.919&60&3 16 00&+45 00 00&''\\
1231&3mm glass&103a&E&1232&1954.742&50&3 17 00&+54 21 00&1st epoch R\\
845&3mm glass&103a&E&845&1953.769&45&3 30 20&+48 19 00&''\\
1249&3mm glass&103a&E&1249&1954.761&50&2 56 30&+48 22 20&''\\
\hline
\end{tabular}
\normalsize
\label{plates}
\end{center}
\end{table*}
%
   \begin{table*}
\begin{center}
      \caption{Schmidt plates used in the Pleiades study.}
\begin{tabular}{cccccccccc}
   \hline
Plate no.&Material&Emulsion&Filter&Field no.&Epoch&Exposure&\multicolumn{2}{c}{Plate Centre}&Notes\\
&&&&&&Time&RA&Dec&\\
&&&&&&(minutes)&\multicolumn{2}{c}{B1950}&\\
\hline
1558&3mm glass&IIIaJ&GG385&418&1987.804&65&3 27 00&+30 00 00&$B_{J}$ Plate\\
1520&3mm glass&IIIaJ&GG385&419&1987.755&60&3 50 00&+30 00 00''\\
4287&3mm glass&IIIaJ&GG385&481&1991.782&55&3 18 00&+25 00 00&''\\
930&3mm glass&IIIaJ&GG385&482&1986.847&60&3 40 00&+25 00 00&''\\
4412&3mm glass&IIIaJ&GG385&483&1992.064&50&4 02 00&+25 00 00&''\\
4887&3mm glass&IIIaJ&GG385&548&1992.763&55&3 30 00&+20 00 00&''\\
315&3mm glass&IIIaJ&GG385&549&1985.957&75&3 51 00&+20 00 00&''\\
4307&3mm glass&IIIaJ&GG385&550&1991.796&50&3 12 00&+20 00 00&''\\
4847&3mm glass&IIIaF&RG610&418&1992.744&90&3 27 00&+30 00 00&2nd Epoch
$R$\\
4859&3mm glass&IIIaF&RG610&419&1992.750&90&3 50 00&+30 00 00&''\\
3618&3mm glass&IIIaF&RG610&481&1990.793&85&3 18 00&+25 00 00&''\\
2992&3mm glass&IIIaF&RG610&482&1989.962&60&3 40 00&+25 00 00&''\\
5612&3mm glass&IIIaF&RG610&483&1993.946&60&4 02 00&+25 00 00&''\\
6405&3mm glass&IIIaF&RG610&548&1995.680&70&3 30 00&+20 00 00&''\\
4290&3mm glass&IIIaF&RG610&549&1991.785&70&3 51 00&+20 00 00&''\\
3010&3mm glass&IIIaF&RG610&550&1989.976&85&3 12 00&+20 00 00&''\\
7017&film&IVN&RG9&418&1996.694&60&3 27 00&+30 00 00&$I$ Plate\\
6526&film&IVN&RG9&419&1995.806&60&3 50 00&+30 00 00&''\\
7008&film&IVN&RG9&481&1996.689&60&3 18 00&+25 00 00&''\\
7022&film&IVN&RG9&482&1996.697&60&3 40 00&+25 00 00&''\\
6025&film&IVN&RG9&483&1994.822&60&4 02 00&+25 00 00&''\\
7026&film&IVN&RG9&548&1996.700&60&3 30 00&+20 00 00&''\\
6474&film&IVN&RG9&549&1995.744&60&3 51 00&+20 00 00&''\\
6444&film&IVN&RG9&550&1995.722&60&3 12 00&+20 00 00&''\\
31&3mm glass&103a&E&31&1949.973&45&4 00 00&+24 13 12&1st Epoch $R$\\
223&3mm glass&103a&E&223&1950.938&45&4 05 28&+18 15 11&''\\
441&3mm glass&103a&E&441&1951.914&50&3 33 36&+24 18 58&''\\
444&3mm glass&103a&E&444&1951.917&50&3 41 24&+18 18 43&''\\
1276&3mm glass&103a&E&1276&1954.894&45&3 07 48&+24 21 54&''\\
1457&3mm glass&103a&E&1457&1955.812&50&3 33 50&+30 18 58&''\\
1460&3mm glass&103a&E&1460&1955.814&50&3 17 24&+18 21 00&''\\
1468&3mm glass&103a&E&1468&1955.861&50&4 00 00&+30 16 01&''\\
\hline
\end{tabular}
\normalsize
\label{PLplates}
\end{center}
\end{table*}
   \begin{figure}
   \resizebox{\hsize}{!}{\includegraphics{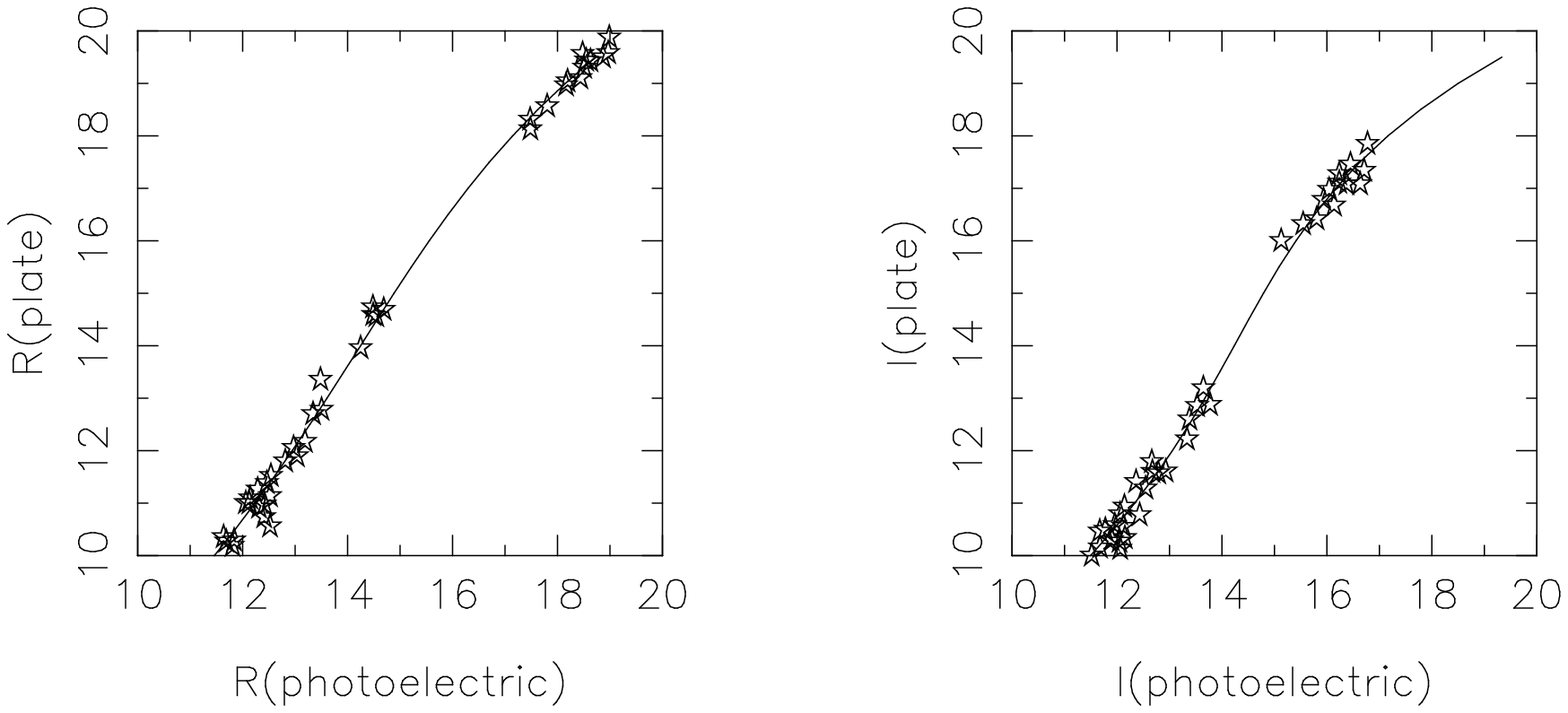}}
      \caption{Comparing the photographic plate magnitudes with the
      measured photoelectric magnitudes for the Alpha Per data
      set. The line shown is a quartic fit to the data. The photoelectric data has been naturalised
      to the photographic passband system.}
         \label{phot1}
   \end{figure}
%

   \begin{figure}
   \resizebox{\hsize}{!}{\includegraphics{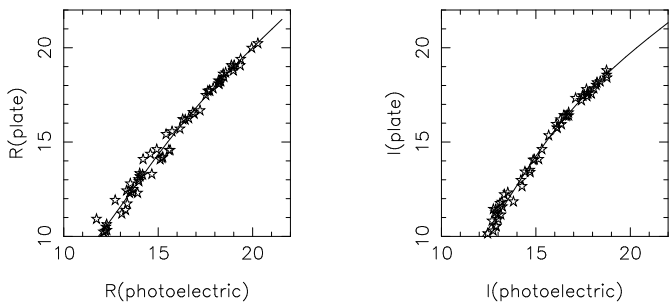}}
      \caption{Comparing the photographic plate magnitudes with the
      measured photoelectric magnitudes for the Pleiades data set. The line shown is a quartic
      fit to the data. Again the photoelectric data has been naturalised
      to the photographic passband system.}
         \label{PLphot1}
   \end{figure}
\subsection{Photometric calibration}
The photometric calibration used by the SuperCOSMOS survey is subject
to large systematic errors especially at the bright end. This
situation can be improved when large numbers of photoelectric
measurements are availible, as is typically the case in open clusters,
using a process of photometric recalibration. To do this
the photographic magnitudes from the sample were compared with
their photoelectric magnitudes. The photoelectric data for
recalibrating the Alpha Per data set was provided by
John Stauffer and with others being taken from Barrado y Navascu\'{e}s et
al~(2003). Figure~\ref{phot1} shows a plot of photoelectric magnitude
vs. photographic, the line shown is a the fourth degree polynomial fit
to the data. The best fit line shown in the graph was then used to
recalibrate the photographic magnitudes. Unfortunately the data from
Barrado y Navascu\'{e}s' only includes stars which fall in one field
(200). Hence the recalibration was carried out on this plate and
propagated through to the other plates. The scatter from the fitted
polynomial was used to estimate the photometric errors to be,
$\sigma_{R} = 0.20$ and  $\sigma_{I} = 0.20$. A similar process was
carried out for the Pleiades using data from Jameson \& Skillen (1989)
and Stauffer (1984). Figure~\ref{PLphot1} shows a plot of photoelectric magnitude
vs. photographic, again the line shown is a the fourth degree polynomial fit
to the data. The photometric errors were found to be, $\sigma_{R} = 0.17$ and  $\sigma_{I} = 0.21$.\\

   \begin{figure}[h!!]
   \resizebox{\hsize}{!}{\includegraphics{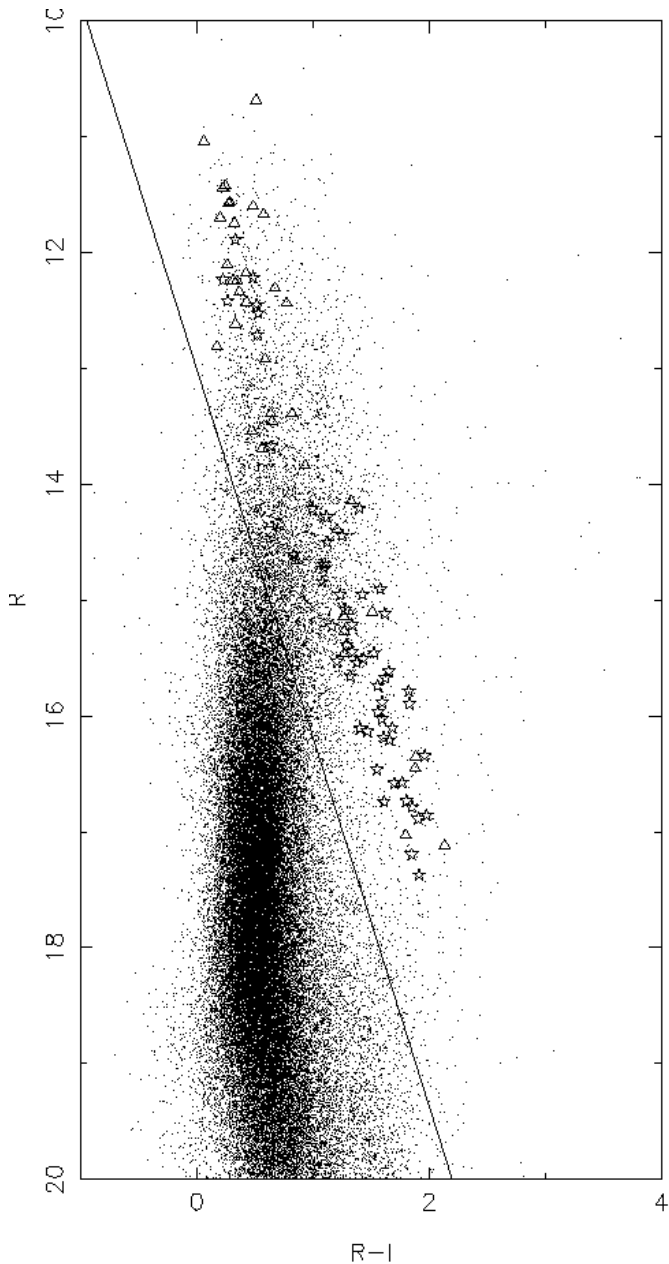}}
      \caption{A colour magnitude diagram for stars in the initial
      sample for Alpha Per. Only stars above the line shown were
      included in the proper motion survey. For symbols see text.}
         \label{cmag}
   \end{figure}
%
   \begin{figure}[h!!]
   \resizebox{\hsize}{!}{\includegraphics{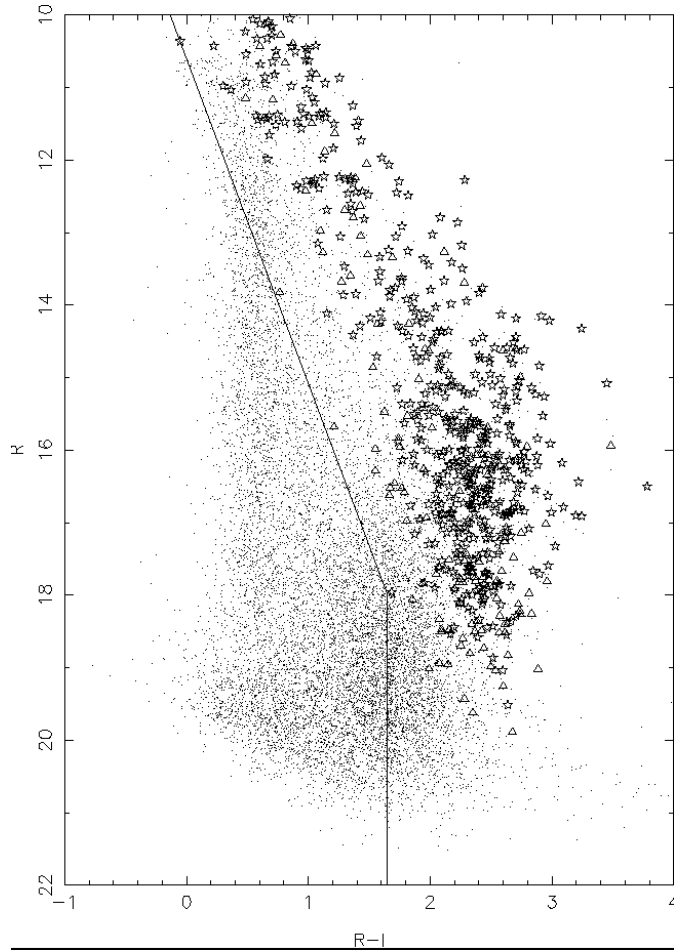}}
      \caption{A colour magnitude diagram for stars in the initial
      sample for the Pleiades. Only stars above the line shown were
      included in the proper motion survey. For symbols see text.}
         \label{PLcmag}
   \end{figure}
\subsection{Colour selection}
After photometric recalibration, data cuts could be made on the
basis of colour and brightness. Firstly it was decided that as stars
brighter than $R=10$ produce high centroiding errors these would be
excluded. In the Alpha Per study the low separation between the
cluster and the field in the Proper Motion Vector Point Diagram
produced a large amount of background star contamination at the faint
end. Hence stars fainter than $R=18$ were excluded. With it's better
separation between the cluster and field such a cut was not necessary
for the Pleiades data set. To reduce the amount of background star
contamination colour selections were made. Figures~\ref{cmag}
and~\ref{PLcmag} show colour magnitude diagrams for each sample. The dots represent stars in the sample while
the star symbols are records from the sample which have been identified as
candidates from both our study and previous studies, the triangles are
star thought to be candidiates by previous studies but not by our study. It can clearly be seen that the
candidate stars from other surveys form a main sequence-like line
distinct from the large mass of background stars. Only stars above the
line shown were used in this study. In the case of the Pleiades
figure~\ref{PLcmag} shows the colour magnitude diagram. The large bulk
of background stars are easily identified and removed. The line shown
is again the colour cut used. Both lines used in the colour cuts were
chosen by inspection
As astrometric errors vary with
magnitude the data was divided into the following magnitude ranges,
for Alpha Per $10<R<12$, $12<R<14$, $14<R<16$ and $16<R<18$ and for
the Pleiades, $10<R<12$, $12<R<14$, $14<R<16$, $16<R<17.5$ and $R>17.5$ . Each of these was treated
independently in the fitting process.\\
\subsection{The fitting process}
The membership probabilities were calculated by a process similar to
that outlined by Sanders (1971). The full mathematical detail is
outlined in Appendix~\ref{math}. To ensure that the data fitting
process is robust a series of twenty sets of simulated data were produced and run
through the fitting program. The data sets were generated by taking uniform random number
distributions. These were converted into gaussian and exponential
distributions with the correct scale lengths, means and variances to
model the expected distributions for the field and cluster stars. Each
data set included 350 cluster stars and 350 field stars, both typical
numbers for the real data. Table~\ref{parerr} shows the results
obtained. It is apparent that there is no significant offset in any of
the calculated parameter values.\\

   \begin{table}
\begin{center}
\begin{tabular}{cccc}
   \hline
Parameter&Input Value&\multicolumn{2}{c}{Fitted Value}\\
&&Mean&Error\\
\hline
$f$&0.5&0.508&0.025\\
$\sigma$&3.5&3.42&0.10\\
$\mu_{xc}$&0.00&-0.01&0.208\\
$\mu_{yc}$&32.0&32.1&0.2\\
$\tau$&25.0&26.5&5.3\\
$\Sigma_{x}$&10.0&10.0&0.4\\
$\mu_{xf}$&0.0&-0.05&0.475\\
\hline
\end{tabular}
\normalsize
\caption{The results of running 20 sets of simulated data through th
eparameter fitting program. With the exception of $f$ all parameters
have units of milliarcseconds per year.}
\label{parerr}
\end{center}
\end{table}
\section{Results}
   \begin{figure}
   \resizebox{\hsize}{!}{\includegraphics{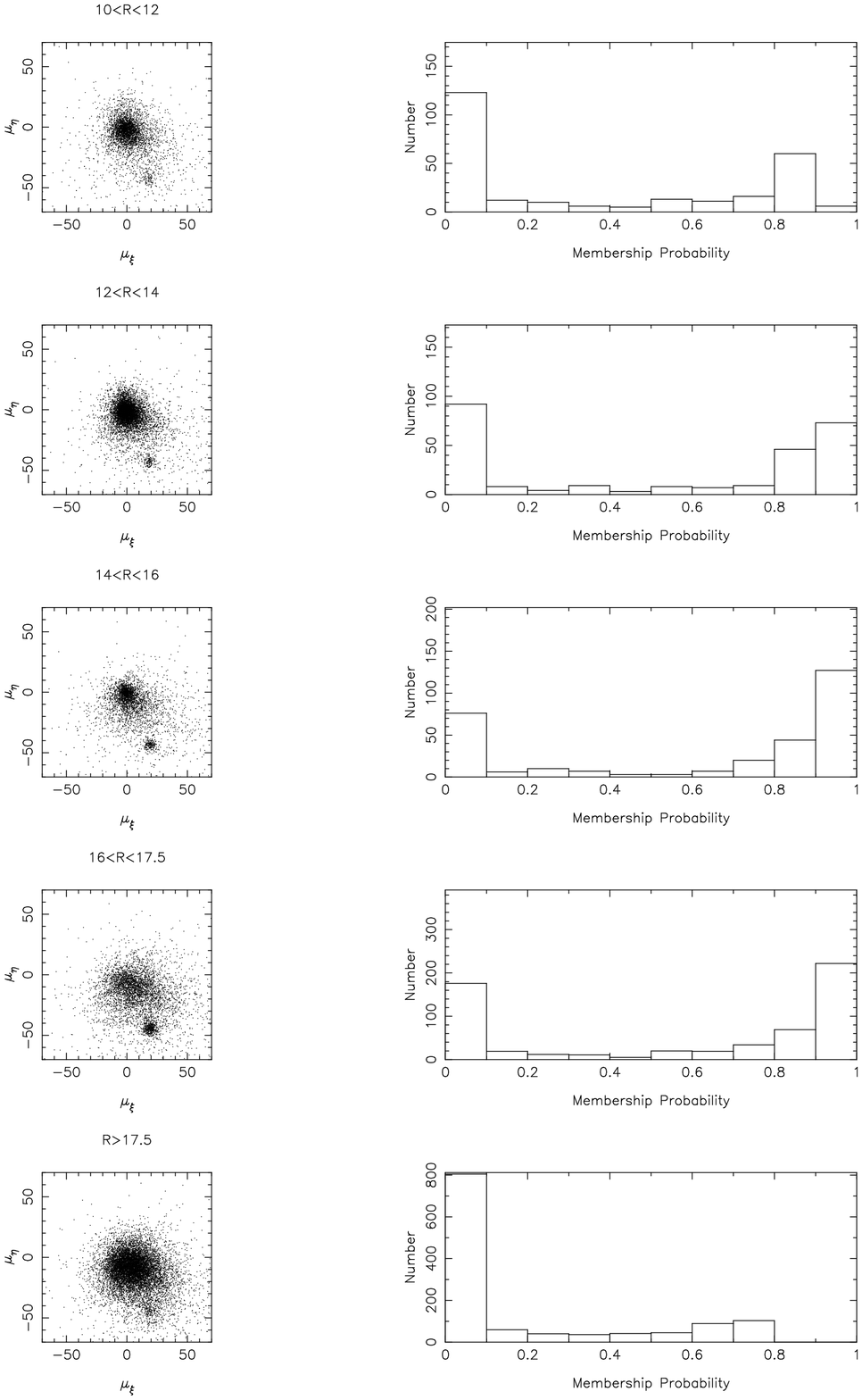}}
      \caption{Proper motion vector point diagrams and probability
      histograms for each magnitude interval in the Pleiades study. In each VPD the cluster
      is the group in the bottom right hand corner, separate from the
      field stars around the origin.}
         \label{prob}
   \end{figure}
%
   \begin{figure}
   \resizebox{\hsize}{!}{\includegraphics{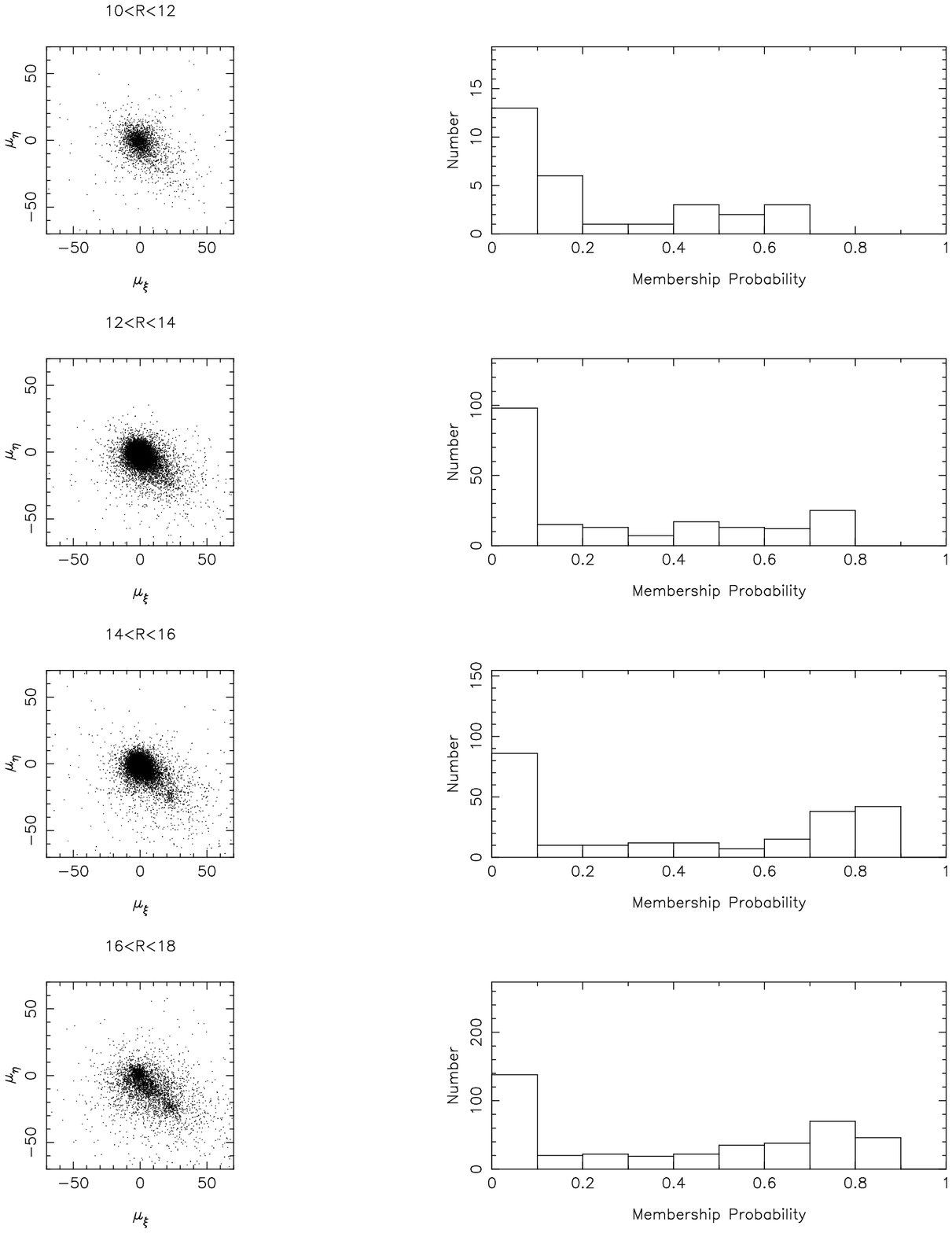}}
      \caption{Proper motion vector point diagrams and probability
      histograms for each magnitude interval in the Alpha Per study. In each VPD the cluster
      is the group in the bottom right hand corner, separate from the
      field stars around the origin.}
         \label{APprob}
   \end{figure}
The proper motion vector point diagrams for each of the magnitude
ranges for both clusters are shown in Figure~\ref{prob} (the Pleiades)
and Figure~\ref{APprob} (Alpha Per) along with probability
histograms. In the case of Alpha Per the cluster is especially apparent in the vector point
diagram $16<R<18$ and consequently it also has the highest number of
high probability cluster members. As the Pleiades study goes far
deeper we can see the cluster becoming more apparent with increasing
magnitude before becoming engulfed by the mass of field stars in the
lowest magnitude interval. Accordingly the number of high probability
members increases with increasing magnitude, the high number of field
stars in the lowest magnitude interval means there are no stars with
membership probabilities $>90\%$. A full list of stars with calculated
membership probabilities greater than $60\%$ is available for both
clusters from CDS in Strasbourg. Example tables are shown in Appendix~\ref{exT}.

\section{Analysis}
   \begin{figure}
   \resizebox{\hsize}{!}{\includegraphics{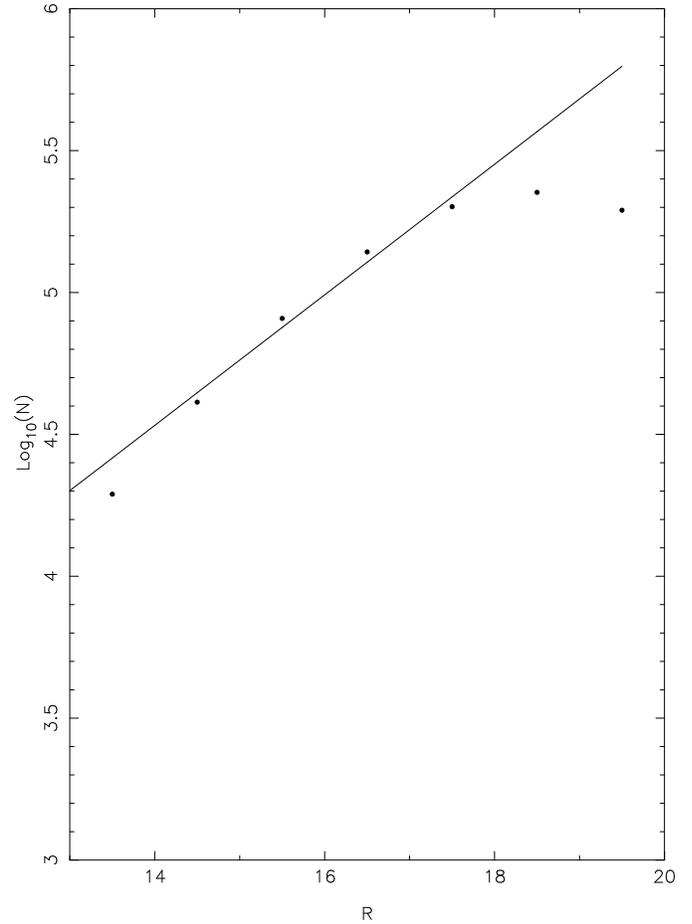}}
      \caption{An estimate of the incompleteness of the study of Alpha
      Per. The
      line shown is a least squares fit.}
         \label{comp1}
   \end{figure}
%
   \begin{figure}
   \resizebox{\hsize}{!}{\includegraphics{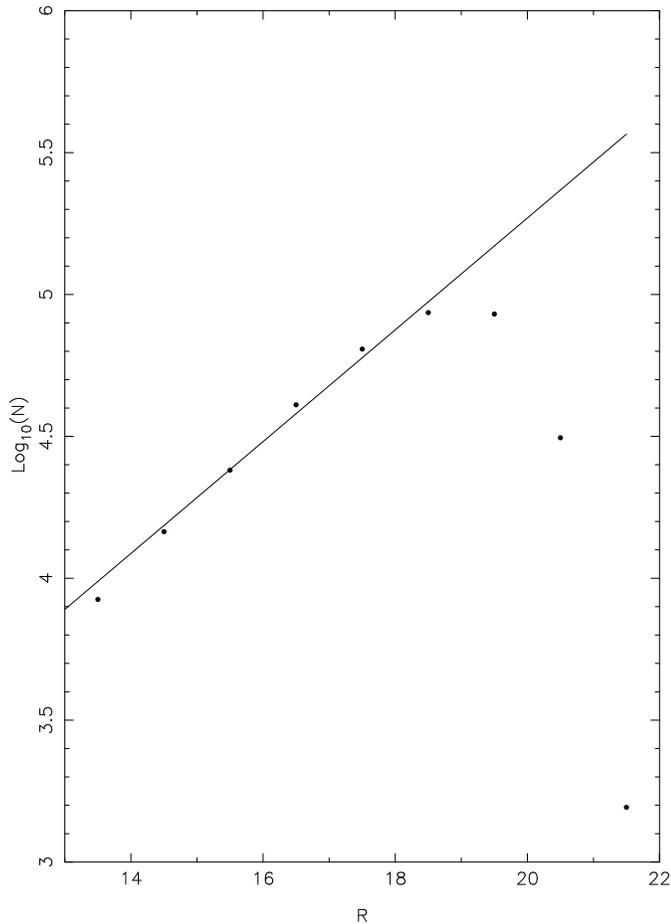}}
      \caption{An estimate of the incompleteness of the study of the Pleiades. The
      line shown is a least squares fit.}
         \label{compPL}
   \end{figure}
%
   \begin{figure}
   \resizebox{\hsize}{!}{\includegraphics{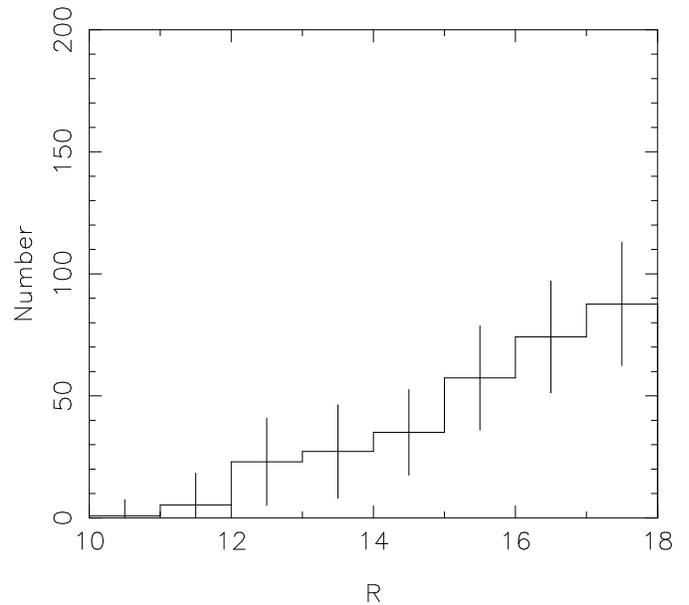}}
      \caption{The Luminosity Function derived for Alpha Per.}
         \label{cLF}
   \end{figure}
%
   \begin{figure}
   \resizebox{\hsize}{!}{\includegraphics{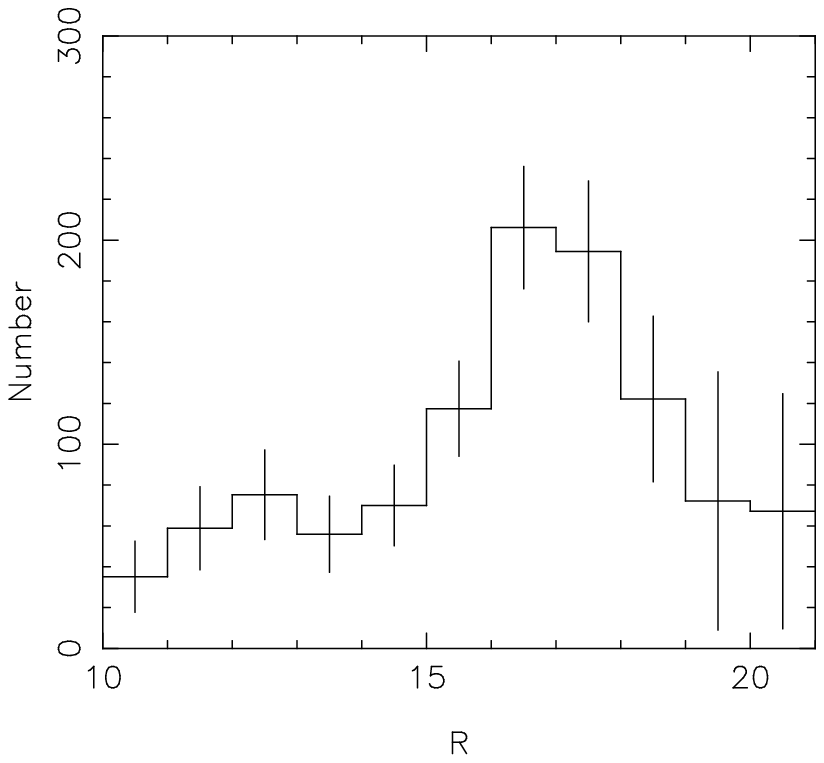}}
      \caption{The Luminosity Function derived for the Pleiades. The
   striking feature is the peak at $R=16-17$}
         \label{PLcLF}
   \end{figure}
\subsection{Luminosity Functions}
As Schmidt plates have poor detection rates close to the plate limit
it was necessary to carry out a completeness estimate. This was done
by measuring how the number of stars in the full stellar sample increased with $R$
magnitude. For a uniform distribution of stars the number in each sample should rise exponentially with increasing
magnitude with any dropoff being caused by incompleteness. An estimate
of this incompleteness can be found by taking the logarithm of the
number of stars in each magnitude interval, fitting a best fit line up
to the point where the dropoff begins and then using the deficit to
estimate the incompleteness. Figure~\ref{comp1} shows such a fit for
the Alpha Per data set and Figure~\ref{compPL} that for the Pleiades sample.\\ 
The luminosity functions were found by summing the membership
probabilities of stars in each one magnitude wide interval. Each
interval was then corrected for incompleteness. Figure~\ref{cLF} shows
the Luminosity Function produced for Alpha Per. Binarity was not taken
into account when producing this. The luminosity function is smooth
and rises with increasing magnitude. The Pleiades luminostiy
function (Figure~\ref{PLcLF}) by contrast has distinct features, the
most obvious of which is the peak at $16<R<17$. The errors shown are
Poisson errors which also take into account errors in the number of
backround stars. 

   \begin{figure}
   \resizebox{\hsize}{!}{\includegraphics{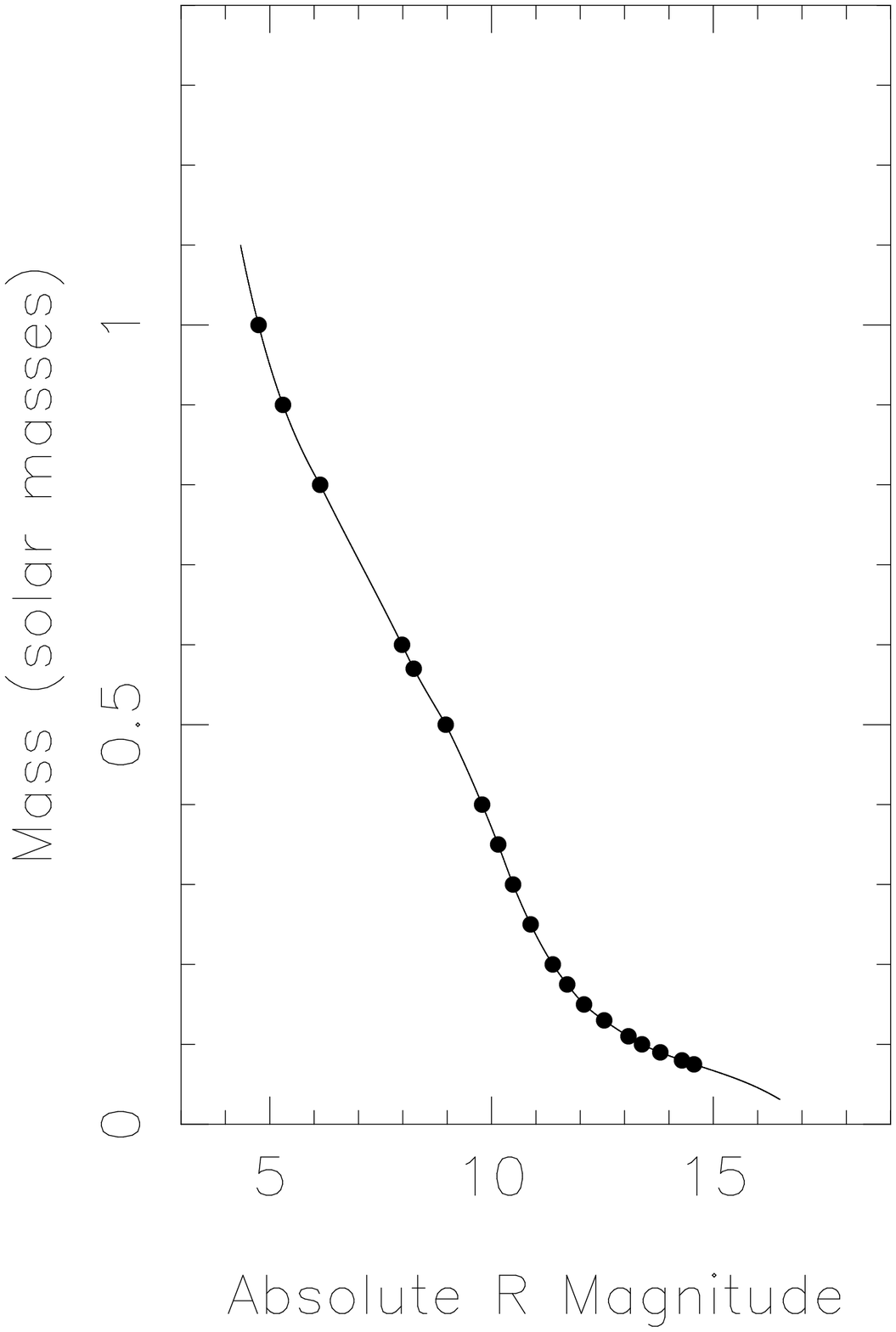}}
      \caption{The mass-luminosity relation used for Alpha Per. The line is a sixth
      degree polynomial fit.}
         \label{APmlplot}

   \end{figure}
   \begin{figure}
   \resizebox{\hsize}{!}{\includegraphics{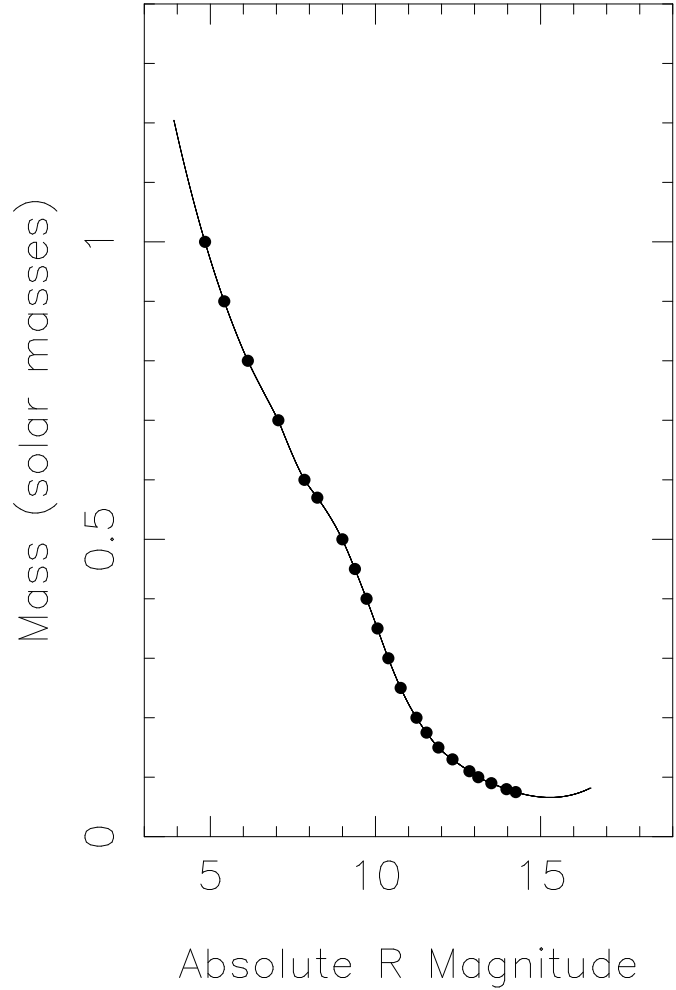}}
      \caption{The mass-luminosity relation used for the Pleiades. The line is a sixth
      degree polynomial fit.}
         \label{mlplot}

   \end{figure}

   \begin{figure}
   \resizebox{\hsize}{!}{\includegraphics{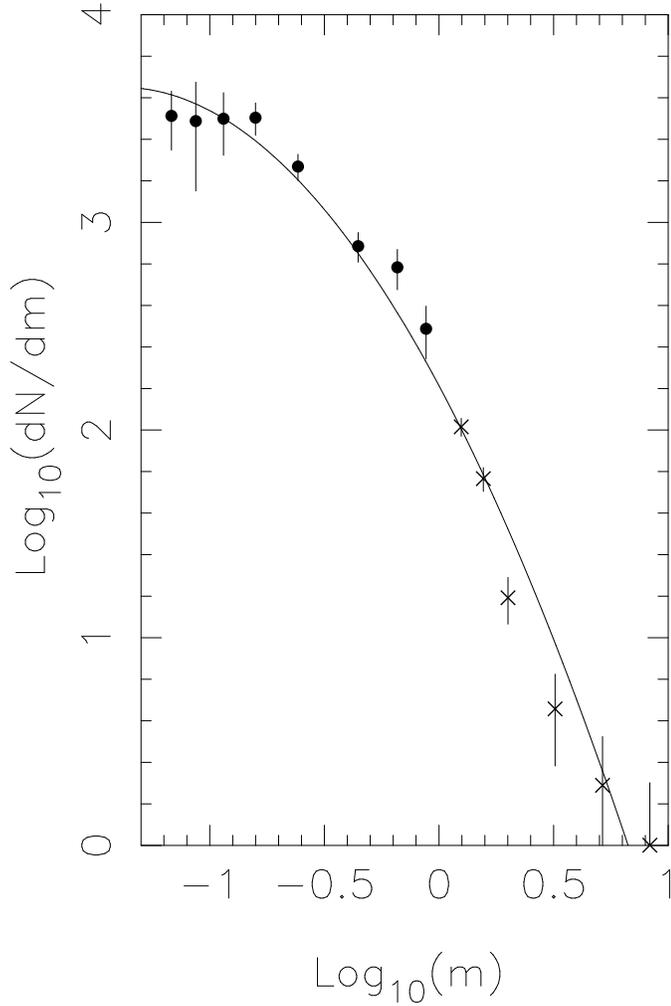}}
      \caption{The derived cluster Mass Function for the Pleiades. Note the
      increasingly shallow gradient towards lower masses. The line
      shown is a log normal fit to the data. The solid circles are
      data points from this survey while the crosses are from Hambly et al
(1999) and refences therein}
         \label{mfall}

   \end{figure}
%
   \begin{figure}
   \resizebox{\hsize}{!}{\includegraphics{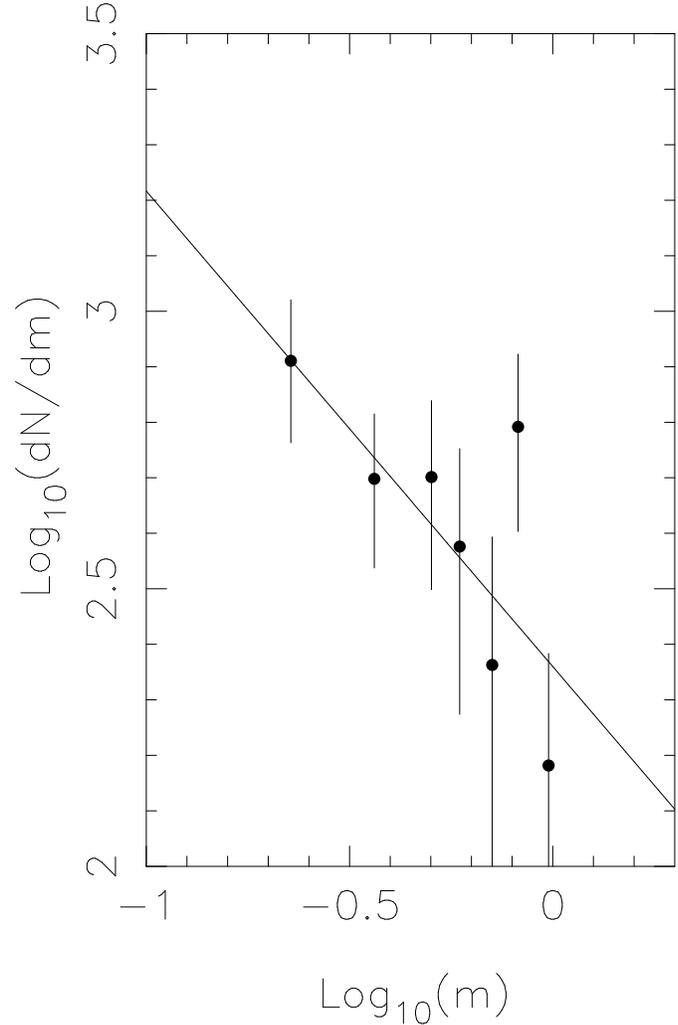}}
      \caption{The derived cluster Mass Function for Alpha Per. The
      line show is a power law fit.}
         \label{APmfall}

   \end{figure}
%
   \begin{figure}
   \resizebox{\hsize}{!}{\includegraphics{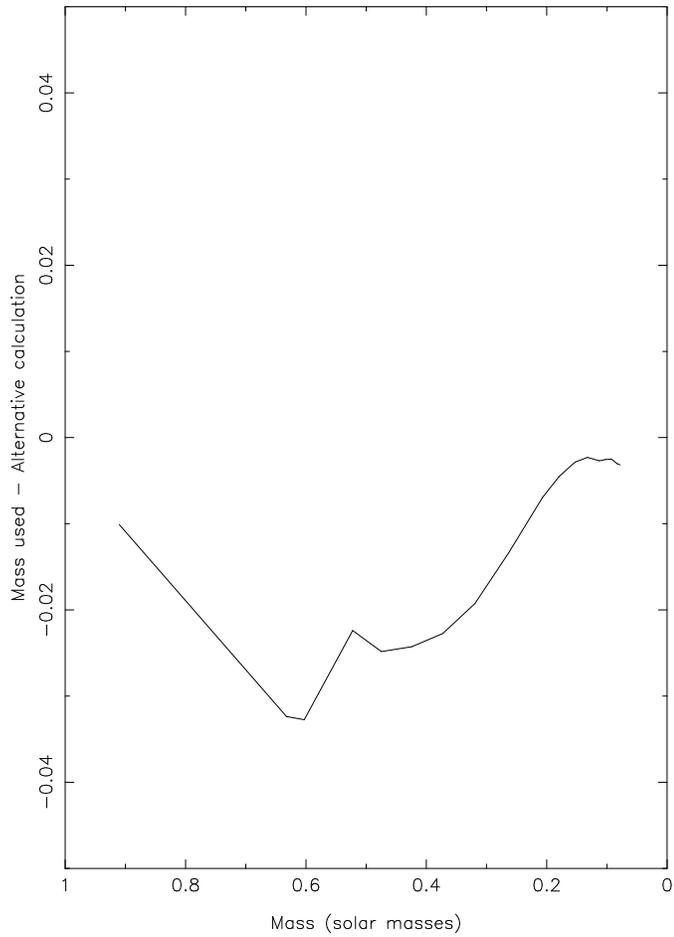}}
      \caption{A plot showing possible systematic errors in the mass
      luminosity relation used for the Pleiades.}
         \label{mlerr}

   \end{figure}
%
   \begin{figure}
   \resizebox{\hsize}{!}{\includegraphics{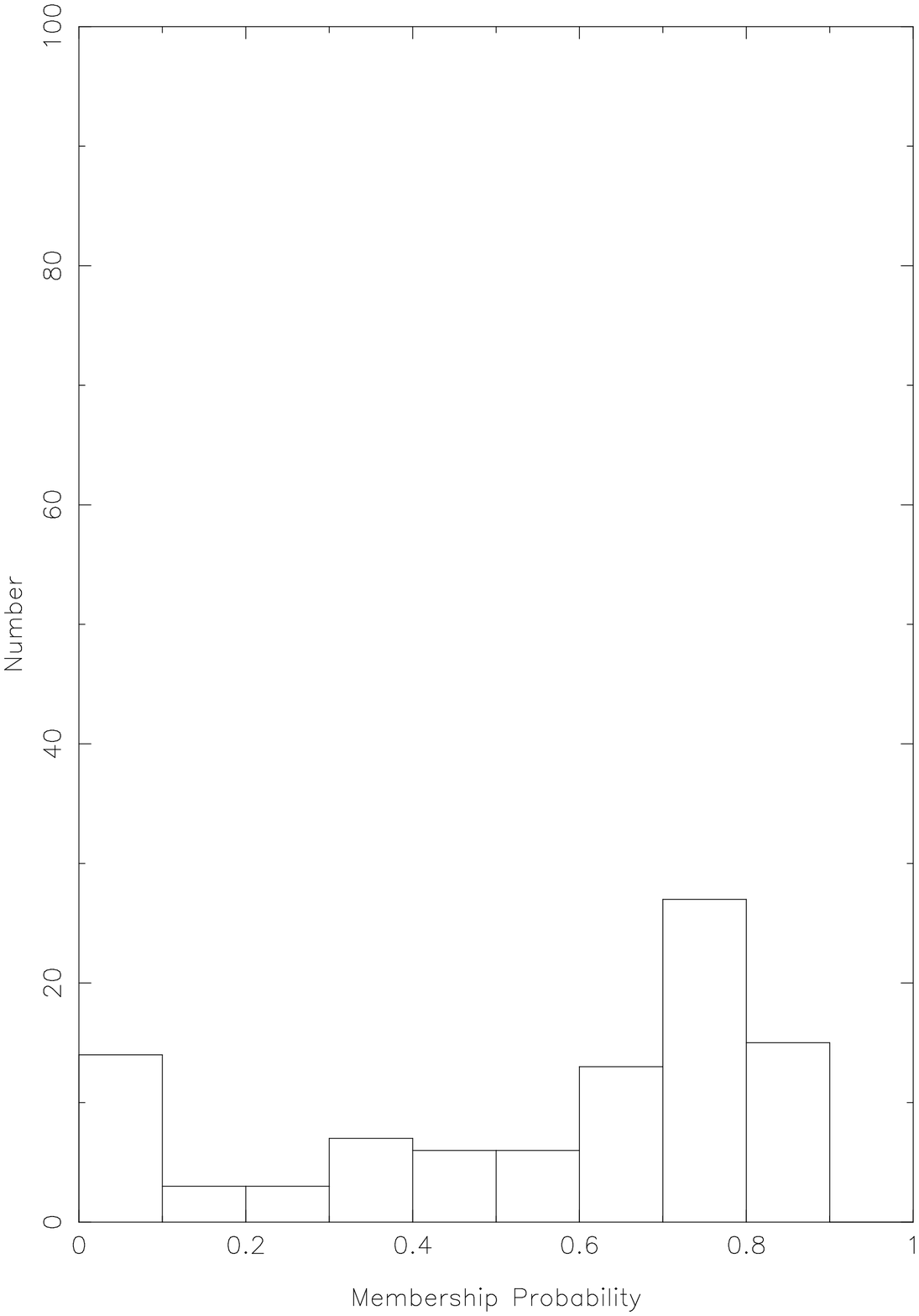}}
      \caption{A histogram of the membership probabilities calculated
      in this study of Alpha Per for stars previously identified as candidate
      members.}
         \label{corl}

   \end{figure}
%
   \begin{figure}
   \resizebox{\hsize}{!}{\includegraphics{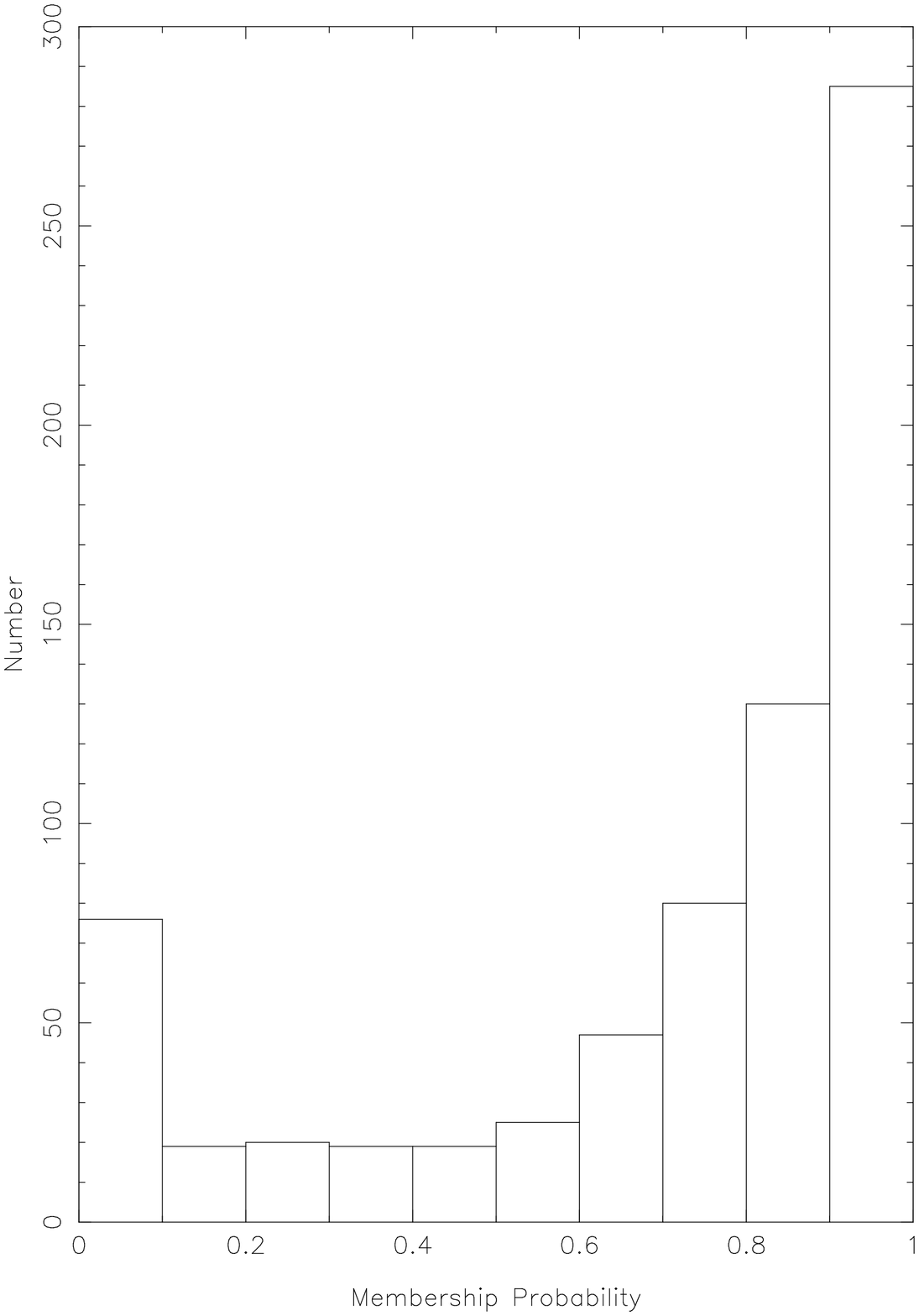}}
      \caption{A histogram of the membership probabilities calculated
      in this study of the Pleiadesfor stars previously identified as candidate
      members.}
         \label{corlPL}

   \end{figure}

\subsection{Mass Function}
To convert the luminosty function to a mass function a mass luminosity
relationship is required. This is provided by the models of Barraffe
et al (1998). Figure~\ref{APmlplot} shows the mass luminosity relation
taken from these models for a 90Myr old population such as Alpha Per. The line shown is a sixth degree
polynomial least squares fit to the data which is used to find the mass of a star of
a particular luminosity. The mass function is found from the
luminosity function, $\Phi(M)$. In some particular magnitude interval of width
$dM$ there will be a number of stars $dN=\Phi(M)dM$. In the
corresponding mass interval the same number $dN$ is given by
$dN=\xi(m)dm$, where $dm$ is the width of the interval and $\xi(m)$ is
the mass function. This can be calculated
by taking the number of stars in a particular luminosity interval and
dividing by the width of that interval in mass. Again binarity was not
taken into account.
The mass function for Alpha Per is shown in Figure~\ref{APmfall}. The mass function
was then fitted by a power law fit of the form,
\begin{equation} 
\xi(m) \propto m^{-\alpha}
\end{equation} 
It was found that $\alpha=0.86^{+0.14}_{-0.19}$. As mentioned earlier Barrado y
Navascu\'{e}s' et al~(2002) derived a mass function for Alpha
Per at lower masses (down to $M=0.035 M_{\odot}$). They also fitted a
power law function of the form, finding that $\alpha=0.59$. It
therefore appears that the mass function flattens off significantly towards lower masses.\\
A mass function was also produced for the Pleiades. This time the mass
luminosity relation was that for a 120 Myr old population and is shown in
Figure~\ref{mlplot}. Again this was taken from Barraffe et al
(1998). The mass function is shown in Figure~\ref{mfall}, the data
points from this survey are supplemented by those from Hambly et al
(1999) and references therein. It has been suggested
by Adams and Fatuzzo~(1996) that the initial stellar mass function can be
approximated by a log normal function as shown below,
\begin{equation}
log_{10}\xi(m) = a_{0} + a_{1}log_{10}m + a_{2}(log_{10}m)^{2} 
\end{equation}
 It is clear that a power law fit is not
appropriate for the mass function derived here; a log normal
was fitted to the Mass function. The parameters are $a_{0}=2.213$,
$a_{1}=-2.069$ and $a_{2}=-0.745$. Hambly et al (1999) produced
a similar log normal fit to the mass function with parameters,
$a_{0}=2.341$,$a_{1}=-2.313$ and $a_{2}=-1.191$. These are in
generally good agreement with the discrepancy in $a_{2}$ probably due
to the poorer constaint of the Hambly study at the faint end. 
The referee has pointed out the possibilty of systematic errors in the
mass-luminosity relationship from the Baraffe et al (1998) models. We
checked for such errors by taking an empirical $BC_{R}$ vs $T_{eff}$
relation derived using the same data and techniques as described in
Hambly et al. (1999) and using it to estimate
the $R$ magnitude from $M_{bol}$ for each of the points taken from
Baraffe's model. The mass of these points was then compared to
that given by the mass-luminosity relation for the recalculated
$R$. The results of this are shown in Figure~\ref{mlerr}.  
\subsection{Comparison with other studies}
As mention earlier there are several other methods to
determine cluster membership. Examining the membership probabilities
of stars from other studies can help both to confirm individual star's
cluster membership and to examine the relevance of both study's
results. It would be worrying if candidates from previous
studies did not have, by and large, high membership probabilities. The candidate stars were paired with the appropriate records
in the catalogue. The maximum pairing radius
was set to 6 arcseconds. A table of those candidate stars for both
clusters with calculated
membership probabilities is availible from CDS in Strasbourg, examples
are provided in Appendix~\ref{exT}. Figure~\ref{corl} shows a histogram of the
membership probabilities for these candidate stars for Alpha Per while
Figure~\ref{corlPL} shows the same graph for the Pleiades study. It is
encouraging that both show that most stars thought to be
members in previous studies have high membership probabilities and
that the large number of probable nonmembers shown in the membership
probability histograms (see Figure~\ref{prob} and Figure~\ref{APprob}) do not appear
here. Both Figure~\ref{corl} and Figure~\ref{corlPL} show a sharp
increase in the number of candidate stars from previous studies at p=60\%. For this reason it was
decided to include only stars with a membership probability greater than this in
the catalogues of probable members for both clusters (available on
CDS). It should be noted that several HHJ objects (see Hambly, Jameson
\& Hawkins 1991) do not have calculated membership probabilities. This
is because these objects appeared highly elliptical on the first epoch
plates SuperCOSMOS measurements and hence were not included in the
present proper motion survey.\\
\section{Conclusion}
We have presented the first proper motion survey of the cluster Alpha
Persei to calculate formal membership probabilities. A  Luminosity Function for the range $R=10$ to
$R=18$ has been produced and a Mass Fuction derived from this. The
Mass Function was fitted with a power law distribution with $\alpha=0.86^{+0.14}_{-0.19}$. A catalogue of 339 high probability members has
been created. This  has been cross-correlated with previous
studies of the cluster showing that a large number of stars previously thought to
be cluster members have high calculated membership
probabilities.
A similar study has also been undertaken on the Pleiades and while not
the first of it's type it has produced several low mass star
candidates and yielded a better constrained Mass Function at the faint end. 
\begin{acknowledgements}
The authors would like to thank John Stauffer for supplying data on
candidate member stars from other surveys and for a prompt and
thorough referee's report.     
\end{acknowledgements}

\clearpage
\begin{appendix}
\section{Mathematics}
\label{math}
In order to calculate membership probabilities, distributions were
fitted to both the cluster stars and the field stars in the vector
point diagram. The cluster stars were fitted with a circularly
symmetric gaussian distribution as in Sanders (1971) however the
field stars were fitted with a decaying exponential in the direction
of cluster proper motion and a gaussian perpendicular to this
direction as in Hambly et al~(1995). In order to fit a
distribution for the field stars the vector point diagram was rotated so that the cluster lay on the
y axis.
The total distribution function $\Phi$ is given by
equation~\ref{phi} where $\Phi_{f}$ is the field star distribution,
$\Phi_{c}$ is the cluster star distribution and $f$ is the
fraction of stars which are field stars. $\Phi_{f}$ is given
in (\ref{phif}) and $\Phi_{c}$ is given
in (\ref{phic}). $c_{o}$ is the normalisation for the exponential between
the limits $\mu_{1}$ and $\mu_{2}$and is given
by (\ref{c}).
\begin{equation}
\Phi = f \Phi_{f} + (1-f) \Phi_{c}
\label{phi}
\end{equation}
\begin{equation}
\Phi_{f} = \frac{c_{o}}{\sqrt{2 \pi} \Sigma_{x}} exp \left(-\frac{(\mu_{x} - \mu_{xf})^{2}}{2
\Sigma_{x}^{2}} - \frac{\mu_{y}}{\tau}\right)
\label{phif}
\end{equation}
\begin{equation}
\Phi_{c} = \frac{1}{2 \pi \sigma^{2}} exp \left(-\frac{(\mu_{x} -
\mu_{xc})^{2}+(\mu_{y} - \mu_{yc})^{2}}{2 \sigma^{2}}\right) 
\label{phic}
\end{equation}
\begin{equation}
c_{o} = \frac{1}{\tau(e^{-\frac{\mu_{2}}{\tau}} - e^{-\frac{\mu_{1}}{\tau}})} 
\label{int}
\end{equation}
\begin{equation}
c_{o} \int_{\mu_{1}}^{\mu_{2}}e^{-\mu_{y}/\tau} d\mu_{y}=1
\label{c}
\end{equation}
 Using the maximum likelihood method 
\begin{equation}
\sum_{i} \frac{\delta ln \Phi_{i}}{\delta \Theta} =0
\label{maxlike}
\end{equation}
(where $\Theta$ is some parameter) the
following set of nonlinear equations are found. 
\begin{equation} 
f: \sum_{i} \frac{\Phi_f - \Phi_c}{\Phi}=0
\label{f}
\end{equation}
\begin{equation} 
\sigma: \sum_{i} \frac{\Phi_{c}}{\Phi}
\left (\frac{(\mu_{x} - \mu_{xc})^{2} + (\mu_{y} - \mu_{yc})^{2}}{\sigma^2}-2\right)=0
\label{sigma}
\end{equation}
\begin{equation}
\Sigma_{x}: \sum_{i} \frac{\Phi_{f}}{\Phi} \left( \frac{(\mu_{x} -
\mu_{xf})^{2}}{\Sigma_{x}^{2}} -1\right)=0
\label{sigmax}
\end{equation}
\begin{equation}
\mu_{xf} : \sum_{i} \frac{\Phi_{f}}{\Phi} (\mu_{x} - \mu_{xf})=0
\label{muxf}
\end{equation}
\begin{equation}
\mu_{xc} : \sum_{i} \frac{\Phi_{c}}{\Phi} (\mu_{x} - \mu_{xc})=0
\label{muxc}
\end{equation}
\begin{equation}
\mu_{yc} : \sum_{i} \frac{\Phi_{c}}{\Phi} (\mu_{y} - \mu_{yc})=0
\label{muyc}
\end{equation}
\begin{equation}
\tau : \sum_{i} \frac{\Phi_{f}}{\Phi}\left( \frac{\mu_{y}}{\tau} - 1 -
c_{o}(\mu_{1}e^{-\frac{\mu_1}{\tau}} -  \mu_{2}e^{-\frac{\mu_2}{\tau}})\right)=0 
\label{tau}
\end{equation}
 These are solved by a
simple bisection algorithm. Applying the calculated values
of the parameters, the probability of the ith star being a cluster
member is,  
\begin{equation}
p_{i}=\frac{(1-f)\Phi_{c i}}{\Phi_{i}} 
\end{equation} 
The fitted values for the parameters for each magnitude interval are
shown in table~\ref{par} for Alpha Per and table~\ref{parPL} for the Pleiades.
\begin{table}[h!!]
\begin{tabular}{cccccccc}
\hline
Interval&$f$&$\sigma$&$\mu_{xc}$&$\mu_{yc}$&$\tau$&$\Sigma_{x}$&$\mu_{xf}$\\
\hline
$10<R<12$&0.94&1.46&0.60&32.4&15.1&13.4&0.39\\
$12<R<14$&0.86&2.42&0.31&33.9&15.0&16.7&2.65\\
$14<R<16$&0.78&2.43&-0.07&33.9&21.8&16.2&1.54\\
$16<R<18$&0.76&2.96&0.72&34.6&20.3&17.1&2.165\\
\hline
\end{tabular}
\caption{The calculated parameters for each magnitude interval for
Alpha Per. With the exception of $f$ all parameters
have units of milliarcsenonds per year.}
\label{par}
\end{table}
\begin{table}[h!!]
\begin{tabular}{cccccccc}
\hline
Interval&$f$&$\sigma$&$\mu_{xc}$&$\mu_{yc}$&$\tau$&$\Sigma_{x}$&$\mu_{xf}$\\
\hline
$10<R<12$&0.77&3.31&-0.11&46.0&17.4&21.2&4.21\\
$12<R<14$&0.72&2.65&0.32&46.8&15.5&20.4&5.86\\
$14<R<16$&0.66&2.56&0.85&47.5&20.0&20.2&8.91\\
$16<R<17.5$&0.63&3.11&0.68&48.1&16.3&20.5&6.92\\
$R>17.5$&0.89&3.52&0.45&48.25&15.1&18.2&6.06\\
\hline
\end{tabular}
\caption{The calculated parameters for each magnitude interval for the
Pleiades. With the exception of $f$ all parameters
have units of milliarcsenonds per year.}
\label{parPL}
\end{table}
\section{Example Tables}
\label{exT}
Example Tables of the catalogue produced. All coordinates are J2000
and all magnitudes are recalibrated photographic magnitudes in the
natural photographic ($R_{59F}$, $I_{IV-N}$) system. Some
membership probabilities may be underestimated at the bright end due
to the low separation of the main sequence from the background stars
at such magnitudes in the colour magnitude diagram.
\begin{table}[h!!]
\begin{tabular}{cccccc}
\hline
No.&RA&Dec&R&I&Membership\\
&&&&&Probability\\
\hline
1 & 2 51 51.84 &+49 1 10.4 &17.55 & 15.66 & 0.8076\\ 
2 & 2 52 0.93 &+48 58 40.6 &16.75 & 15.25 & 0.6804\\ 
3 & 2 53 22.99 &+48 12 22.2 &11.55 & 11.31 & 0.6251\\ 
4 & 2 53 45.30 &+47 26 3.5 &14.49 & 13.76 & 0.6089\\ 
5 & 2 54 12.10 &+47 56 54.8 &15.74 & 14.65 & 0.6006\\ 
\hline
\end{tabular}
\caption{An example table of the the catalogue of high probability
members of Alpha Per.}
\label{ExAP}
\end{table}

\begin{table}[h!!]
\begin{tabular}{ccc}
\hline
Name&Membership&No.\\
&Probability&\\
\hline
HE 767  & 0.000001 & \\ 
HE 389  & 0.005409 & \\ 
AP225   & 0.354993 & \\ 
AP121   & 0.590439 & \\ 
AP102   & 0.000001 & \\ 
AP139   & 0.497886 & \\ 
AP 41   & 0.115724 & \\ 
AP118   & 0.007252 & \\
AP264   & 0.000004 & \\ 
AP 25   & 0.687682 & 149\\
\hline
\end{tabular}
\caption{An example table of the the catalogue of candidate members of
Alpha Per from other studies with calculated membership
probabilities. Several of these stars have alternative names, for
comprehensive cross-identifications, see the Stauffer and Prosser
Catalogue of open cluster data (Stauffer, private communication).}
\label{ExAPCross}
\end{table}

\begin{table}[h!!]
\begin{tabular}{cccccc}
\hline
No.&RA&Dec&R&I&Membership\\
&&&&&Probability\\
\hline
1 & 3 27 35.58 &+24 31 43.6 &12.767271 & 12.245660 & 0.926547\\ 
2 & 3 27 37.75 &+24 59 0.8 &20.631550 & 19.037104 & 0.760211\\ 
3 & 3 27 54.24 &+24 56 11.7 &16.532923 & 14.738454 & 0.943296\\ 
4 & 3 28 1.54 &+23 4 43.0 &17.342480 & 15.450493 & 0.958140\\ 
5 & 3 28 14.3 &+26 20 35.3 &13.848833 & 13.519602 & 0.639352\\
\hline
\end{tabular}
\caption{An example table of the the catalogue of high probability
members of the Pleiades.}
\label{ExPL}
\end{table}
\begin{table}[h!!]
\begin{tabular}{ccc}
\hline
Name&Membership&No.\\
&Probability&\\
\hline
hcg 2  & 0.956371 & 58\\ 
hcg 6  & 0.933995 & 66\\ 
hcg 11 & 0.954879 & 92\\
hcg 12 & 0.904731 & 95\\ 
hcg 13 & 0.690341 & 96\\ 
hcg 16 & 0.906718 & 99\\
\hline
\end{tabular}
\caption{An example table of the the catalogue of candidate members of
the Pleiades from other studies with calculated membership
probabilities. Several of these stars have alternative names, see the Stauffer and Prosser
Catalogue of open cluster data (Stauffer, private communication).}
\label{ExPLCross}
\end{table}
\end{appendix}
\end{document}